\documentclass[doublecol,linenumbers]{epl2} 
\usepackage{amssymb}   
\usepackage{amsmath}
\title{Conjugate coupling induced symmetry breaking and quenched oscillations }
\shorttitle{Conjugate coupling induced symmetry breaking and quenched oscillations} 

\author{ K. Ponrasu\inst{1}\!\!\!,  K. Sathiyadevi\inst{1}\!\!\!,   V. K. Chandrasekar\inst{1} \and M. Lakshmanan\inst{2}}
\shortauthor{K. Ponrasu \etal}

\institute{                    
	\inst{1} Centre for Nonlinear Science \& Engineering, School of Electrical \& Electronics Engineering, \\SASTRA Deemed University, Thanjavur -613 401, Tamil Nadu, India. \\
	\inst{2} Centre for Nonlinear Dynamics, School of Physics, Bharathidasan University, Tiruchirappalli - 620 024,\\ Tamil Nadu, India. }

\abstract{ Spontaneous symmetry breaking (SSB) is essential and plays a vital role many natural phenomena,  including formation of Turing pattern in organisms and complex patterns in brain dynamics.  In this work,  we investigate whether  a  set of coupled Stuart-Landau oscillators can exhibit spontaneous symmetry breaking when the oscillators are interacting through dissimilar variables or conjugate coupling.  We find  the emergence of SSB state with   coexisting  distinct dynamical states  in the  parametric space  and   show  how the system transits  from symmetry breaking state to out-of-phase synchronized (OPS) state while  admitting multistabilities among the dynamical states.  Further, we also investigate the effect of feedback factor on  SSB as well as   oscillation quenching  states and we  point out that the decreasing feedback factor  completely suppresses SSB  and oscillation death states.  	 Interestingly, we also find the feedback factor completely diminishes only symmetry breaking oscillation and  oscillation  death (OD) states  but it does not affect the nontrivial amplitude death (NAD) state.    Finally, we have deduced the analytical stability conditions  for  in-phase and out-of-phase oscillations, as well as  amplitude  and oscillation death states. 
}
\begin{document}
	\maketitle
	\section{Introduction }
	
\par 	Coupled nonlinear oscillatory systems exhibit distinct collective dynamical behaviors including various synchronization patterns \cite{book1}-\cite{syn3}, and oscillation quenching states \cite{os1}-\cite{AD-OD}, which have been observed  in physical, chemical and biological systems \cite{ap1}-\cite{ap3}.  Among them, one of the intriguing phenomena  observed recently is the symmetry breaking states which  play a crucial role in many natural systems.  Importantly, in biological systems the spontaneous  symmetry breaking (SSB) is essential for cell movement, developmental patterning of  vertebrates and also for  Turing pattern formation in organism\cite{ssb_turing}-\cite{turing}.  In the neural network, SSB leads to the formation of diverse complex patterns \cite{comp}.	Due to the above features, the mechanism underlying such SSB and the situation in which such SSB occurs are of great interest.  Recently, the trade-off between the attractive and repulsive coupling induced SSB has	been reported \cite{sbo}.    
	
	\par  Oscillation quenching is another emergent intriguing phenomenon  found in  various  chemical and biological systems   \cite{od_chem_osc}-\cite{liu_od}.  The  quenching of  oscillation is  further subdivided into two categories, i.e. (i) amplitude death (AD) and (ii) oscillation death (OD).  The AD exists when the coupled oscillators interact in such a way as to suppress each other's oscillations and collectively attain a  homogeneous steady state \cite{saxena}. In the earlier reports, it has been reported that three main factors can cause AD, namely  (i) coupling through dissimilar variables \cite{conj_td}, (ii) time delay \cite{td0}-{\cite{td3}} and (iii) frequency mismatch of the oscillators \cite{fm}. Later, such dynamical behavior was also observed in dynamic coupling \cite{dynamic}, direct-indirect  coupling \cite{dir_indi} and even in mean field diffusive coupling \cite{ADOD}.   In many practical situations the nontrivial AD is a desirable control mechanism to suppress harmful oscillations in laser systems and neuronal activity, where a constant output is needed and fluctuations should be suppressed \cite{laser_ad}-\cite{neu_ad}.  Further, the coupled oscillators can  exhibit another type of oscillation quenching phenomenon, namely oscillation death (OD) where the strong interaction among the  oscillators  leads the system to  attain an inhomogeneous steady state (IHSS) \cite{revi2},\cite{od_prl}. The OD can occur under  distinct coupling schemes such as  time-delayed coupling  \cite{delay_od1}-\cite{delay_od2} conjugate coupling \cite{conj1},  mean field diffusive\cite{ADOD} and repulsive interactions \cite{rep_od}.  The OD state exhibits significant implications in many biological systems such as synthetic genetic networks\cite{sgo1}-\cite{sgo2}, cellular differentiation \cite{stem} and cardiovascular phenomena\cite{cardio}.  Even though the OD state has many practical applications, in some situations the oscillation death is detrimental and   should be avoided.    Later, many investigations have been carried out to  control oscillation death or revival of oscillation from death and several authors have    reported the  ubiquitous phenomenon of feedback  as a mechanism to control death state  \cite{liu_revive}-{\cite{AD_revok}}.

\par 	Moreover, many investigations have been accomplished  to study the dynamical behavior in coupled oscillators where  the coupling among the oscillations is established through similar variables. On the other hand, in many real situations  coupling via dissimilar variable is also desirable and such a coupling  is known as conjugate coupling.   In a variety of experimental situations the coupling via a  conjugate variable is essential where  the subsystems are coupled by feeding the output of one into the other.  For example, in laser experiments, \cite{conj_scl}  the light emitted from each laser diode is monitored by a photo-diode detector (whose	AC signal is amplified or reduced) by feed back to the other. Further, in the ecological context,  the dynamics of two predator-prey systems  that are coupled via cross predation is also investigated, in which each predator consumes prey in the 	other system which leads to stabilizing the food web to a new equilibrium \cite{prey}.	In addition,    theoretical predictions reveal that time delay is not necessary,  and coupling with dissimilar variables can lead to AD even when the coupling is bidirectional and diffusive \cite{conj_td}. Further, the emergence of  a large number of states  such as homogeneous and inhomogeneous  oscillation death states and homogeneous oscillatory states  have been found in a  ring of identical Stuart-Landau oscillators where each oscillator interacts with its nearest neighbors via a diffusive conjugate coupling \cite{conj1}. Recently, shrinking of AD regions were reported by adjusting the asymmetry factors in conjugately coupled systems \cite{conj(2017)} and the emergence of  explosive death  has also been found in a network of coupled  van der Pol oscillators while coupling via conjugate variables \cite{conj_vdp(2018)}. 
	
\par Motivated by the above studies, in this letter, we study the  dynamical behavior in a system of  conjugately coupled Stuart-Landau oscillators. We investigate,  \textit{whether the conjugate coupling can be able to induce spontaneous symmetry breaking state.}  Interestingly,  we find that the conjugate coupling among the oscillators breaks the system symmetry spontaneously and gives rise to SSB state along with distinct multistable regions. Through a study of the associated  basin of attraction \revision{\cite{basin}}, we show how the transition occurs  from SSB state  to OPS state. In particular, the transition to OPS state takes place via two distinct types of transitions: (i) sudden transition from SSB state  to OPS state and (ii) continuous transition (i.e. increasing parameter increases the OPS basin while decreasing the SSB basin) to OPS state.    In addition, we  analyze the effect of feedback  by introducing the feedback factor in the system and  show that  decreasing feedback factor suppresses the symmetry breaking oscillations.  Moreover, we find the mean field conjugate coupling can induce the oscillation death (OD) state which coexists with nontrivial amplitude death (NAD) state. The impact of feedback factor is also investigated on both OD and NAD states.   The observed oscillation death state shrinks while decreasing the feedback factor.  The feedback factor only suppresses oscillation death but it does not affect the nontrivial amplitude death state.  
	This type of dynamical behavior can be useful and have potential applications  in laser systems where we need a constant output  and   the fluctuation should be shorted out.

	\section{Model} 
	In order to exemplify our results, we consider a general, paradigmatic model of Stuart-Landau  limit cycle oscillators \cite{model_stu},  which are coupled through mean-field of the dissimilar variables. The governing equations can be given by 
	
	\begin{eqnarray}
	\dot{z_k} = f(z_k)+\epsilon [{\mathrm{Im}}{( \mathbf Z)}-a\, {\mathrm{Re}}{(z_k)}], \quad {k=1,2}.
	\label{model2}
	\end{eqnarray}
	where $f(z_k) =(\lambda+i\omega-|z_k|^2)z_k$ and $z_k=re^{i\phi_k} = x_k+iy_k  \in C$, (${k=1,2}$).   $x_k$ and $y_k$ are the state variables of the system. $ \mathbf Z = \frac{z_1+z_2}{2}$,  is the mean field of the coupled system.     {The intrinsic parameter  of the system is denoted by $\lambda$ which corresponds to  the distance from the Hopf bifurcation point \cite{lambda}}.  $\omega$ is the natural frequency,  $\epsilon$ be the coupling strength and $a$ $(0 \le a \le 1)$ is the feedback control parameter.   To solve  Eq. (\ref{model2}) we use Runge-Kutta fourth order method with  time step  value of 0.01.
	\begin{figure*}
		\centering
		\hspace{-0.1cm}
		\includegraphics[width=18.00cm]{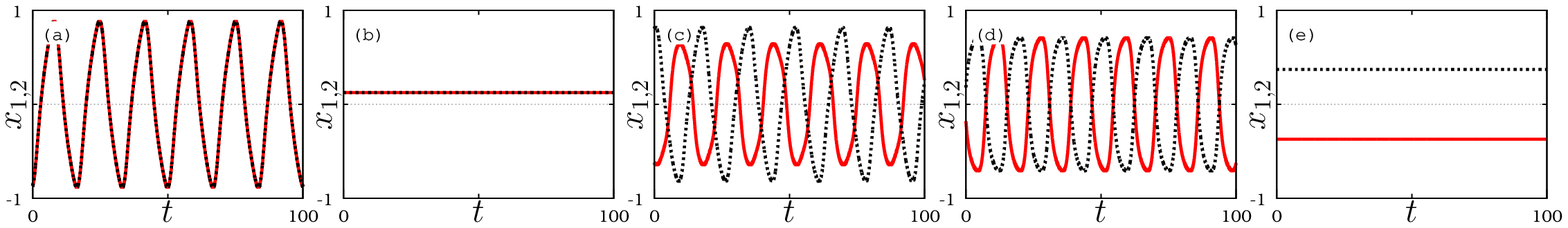}
		\caption{Temporal dynamics of the collective system (\ref{model2}) as a function of coupling strength $\epsilon$. (a)  In-phase synchronized oscillation (IPS) for  $\epsilon=0.2$, (b)  nontrivial amplitude death (NAD) state for  $\epsilon=0.4$, (c)    symmetry breaking oscillation (SSB) for $\epsilon=0.5$, (d)   out-phase synchronized oscillation (OPS) state  for  $\epsilon=0.6$  and  (e)  oscillation death (OD) state for $\epsilon=0.9$.  Other parameters: $\omega=0.5$, $a=1.0$ and  $\lambda=1.0$.}
		\label{tim}
	\end{figure*}

	\section{Dynamical behavior and  symmetry breaking dynamics}
	At first, the dynamical behavior of the collective system is  inspected  through time series  for a  fixed value of natural frequency $\omega=0.5$ (for fixed values of $\lambda = 1.0$ and $a =1.0$)  and  by varying the  coupling strength $\epsilon$ in Fig.~\ref{tim}.  Emergence of in-phase synchronized (IPS) oscillations  is  observed at  $\epsilon=0.2$ as shown in Fig. \ref{tim}(a). In this state, both the oscillators oscillate with the  same amplitude and phase. On increasing the coupling strength to $\epsilon=0.4$, both the oscillators attain a  nontrivial homogeneous steady state or amplitude death (NAD) state (see Fig.~\ref{tim}(b)). Fig.~\ref{tim}(c) depicts that at a  critical value of the coupling strength $\epsilon=0.5$, the system exhibits a  spontaneous symmetry breaking (SSB) state, where both the oscillators oscillate with different amplitudes.  It is due to spontaneous breaking of the permutational or translational symmetry $(z_1 \pm z_2)$ of the system. Further, increasing of the coupling  strength  to $\epsilon=0.6$, one can note that the oscillators adjust their amplitudes so as to  give rise out-of-phase synchronized (OPS) oscillations as shown in Fig. \ref{tim}(d).   Increasing the coupling strength to  a higher value  $\epsilon=0.9$  leads to oscillation death (OD) state where both the oscillators attain inhomogeneous steady states (IHSS)  ( see Fig. \ref{tim}(e)).

\begin{figure}[h]
	\centering
	\vspace{-0.5cm}
	\includegraphics[width=8.0cm]{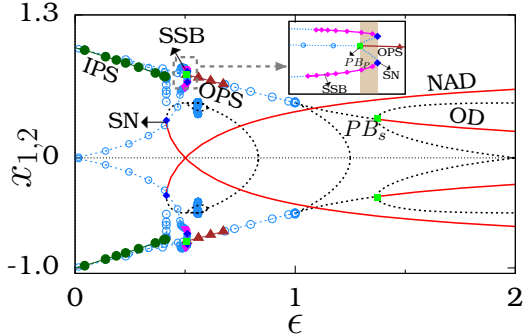}
		\vspace{-0.5cm}
	\caption{  Bifurcation diagram for a pair  of conjugately  coupled  Stuart-Landau oscillators for  $\omega=0.5$ and the enlarged image of  dashed lines (grey)  is shown in the inset.   In the figure  dotted (black) lines and solid (red) line represent the unstable and stable oscillation death states. Filled circles  and triangles denote  stable in-phase synchronized  and  out-of-phase synchronized states, respectively. Spontaneous symmetry breaking states are  denoted by diamond points (pink) and the unfilled circles represent unstable limit cycle oscillations. $SN$, $PB_P$ and $PB_s$  are  the saddle-node, pitchfork bifurcation of periodic orbits and subcritical pitchfork bifurcation points, respectively. Fig. 2 is also plotted along the line $L_1$  in Fig. 3.  Other parameters are same as in Fig.~\ref{tim}.}
	\label{swing}
	\vspace{-0.2cm}
\end{figure} 

	In addition, to understand  the dynamical transitions and multistability of the considered system, a bifurcation analysis has been carried out by fixing the natural frequency at  $\omega=0.5$. The corresponding bifurcation diagrams are illustrated in Fig~\ref{swing} (obtained using XPPAUT software).   At this  frequency  $\omega=0.5$, the systems exhibit IPS state for lower values of coupling strength $\epsilon$ (see Fig. \ref{swing}) and it is evident that increasing $\epsilon$ destabilizes the IPS state and gives birth to  nontrivial amplitude death (NAD) state. The observed NAD state  is stable for further entire  range of coupling strength.  In addition, the unstable out-of-phase synchronized (OPS) state is coexistant with stable IPS state and gets stabilized via pitchfork bifurcation of periodic orbit $(PB_P)$ which is clearly depicted in the inset of Fig.~\ref{swing}. Before stabilization of OPS state, we find that there arises spontaneous symmetry breaking (SSB) state due to  the breaking up of the permutational or translational symmetry $(z_1 \pm z_2)$ of the system.  The OPS state is further destabilized while increasing $\epsilon$, which gives birth to oscillation death (OD) state at strong values of $\epsilon$ via subcritical pitchfork bifurcation ($PB_s$).  Additionally, the coexistence of NAD state is observed in the OD region (see Fig. \ref{swing}).  Besides,  the bifurcation analysis indicates that  both the oscillators in the IPS state (i.e. both the oscillators oscillate with same amplitude and phase) transit to nontrivial AD state (i.e. the oscillators  attain the homogeneous  steady state with nontrivial values) on increasing the strength of coupling. On the other hand, the OPS state (oscillations with same amplitude and anti-phase) emerges from the SSB (oscillators oscillate with different amplitudes) state and transit to  oscillation death state on increasing the coupling strength. The above detailed bifurcation analysis was carried out along the line $L_1$ in  Fig. \ref{glob}(a).  The inset in   Fig.~\ref{swing} illustrates the enlarged image of  the dashed lines (grey), which delineates the  dynamical transition in the symmetry breaking regions.  Before the stabilization of the OPS state, the emergence of stable SSB  state is  evident  from the inset of  Fig.~\ref{swing}, which is further   destabilized via saddle-node (SN) bifurcation. In addition, as noted before we find that  the OPS state is stabilized via pitchfork bifurcation of periodic orbits ($PB_P$).   The shaded region in  the inset denotes the coexistence of OPS and SSB states. Comparing Fig.~\ref{swing}  and the inset, one can note that the shaded region denotes the tristabilty region where SSB, OPS and NAD states are stable. The global dynamical behavior of  Eq.~(\ref{model2}) is further detailed in the following section.

	\section {Global behavior of the coupled system}
		\begin{figure}[h]
			\centering
			\hspace{-0.2cm}
			\includegraphics[width=8.9cm]{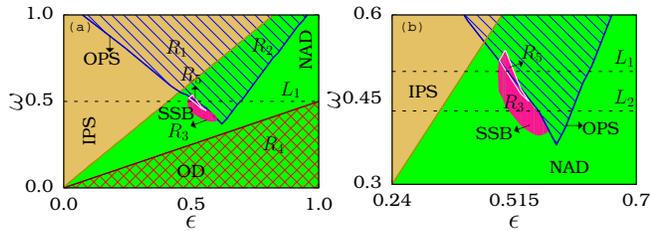}
			\vspace{-0.4cm}
			\caption{(a) Two parameter phase diagram in ($\epsilon, \omega$) space.  (b) Enlarged image of spontaneous symmetry breaking region.  IPS and OPS represent stable in-phase synchronized and out-of-phase synchronized regions. NAD and OD are the nontrivial amplitude death and oscillation death states. Spontaneous  symmetry breaking state is  denoted by SSB.  $R_1$, $R_2$, $R_3$ and $R_4$ represent the bistability regions of IPS-OPS, OPS-NAD, SSB-NAD and NAD-OD, respectively. The tristability region of SSB-NAD-OPS is represented by $R_5$.   The boundary lines of each state denote the analytically obtained stability curves.   Other parameters are $\lambda=1.0$, $a=1.0$.    }
			\label{glob}
			\vspace{-0.2cm}
		\end{figure} 	
	
	In order to  understand the global dynamical behavior of the  system (\ref{model2}), a two parameter phase diagram is constructed in $(\epsilon,\omega)$ space  as shown in Fig. \ref{glob}. From Fig. \ref{glob}(a), at lower values of  $\omega$, for  increasing coupling we have observed  stable IPS  states at very low ranges of $\epsilon$, and increment of $\epsilon$ further gives rise to NAD state for  all values of coupling strength.  In this case, the IPS state loses its stability and ultimately  the system transits  to  stable  NAD state.   Increase of  coupling strength also exhibits bistability  between OD and NAD states.    At lower values of  coupling strength, on  increasing the natural frequency the system transits from monostable nature of IPS state to bistable state. In other words, coexistence of  IPS and OPS states can be observed.  It is also noticed that at critical values of $\epsilon$ and $\omega$, the system exhibits SSB.  In this state the transition  to OPS state takes place through symmetry breaking oscillations and the NAD and OPS states also coexist with SSB. At strong values of  $\epsilon$ and $\omega$ the system admits  only NAD state.   The regions $R_1$, $R_2$, $R_3$ and $R_4$ are the bistability regions between IPS-OPS, OPS-AD, SSB-NAD and NAD-OD, respectively.  Further, to inspect the dynamical transitions in the  symmetry breaking region, we enlarge the symmetry breaking region in Fig. \ref{glob}(b) where we find two distinct types of transitions along the lines $L_1$ and $L_2$ which  are illustrated  through a study of basin of attraction in Fig.~\ref{basin}.   Interestingly, in the SSB region we find the  coexistence  of  three stable states (i.e. tristability), namely SSB, OPS and NAD,  which is denoted as   $R_5$ in Fig. \ref{glob}(b).  Additionally, we have  investigated how the transition takes place from the SSB to OPS state  in the following section. 
	\section{Dynamical transition through basin of attraction in SSB region}
		\begin{figure}[h]
			\centering
			\hspace{-0.4cm}
			\includegraphics[width=8.0cm]{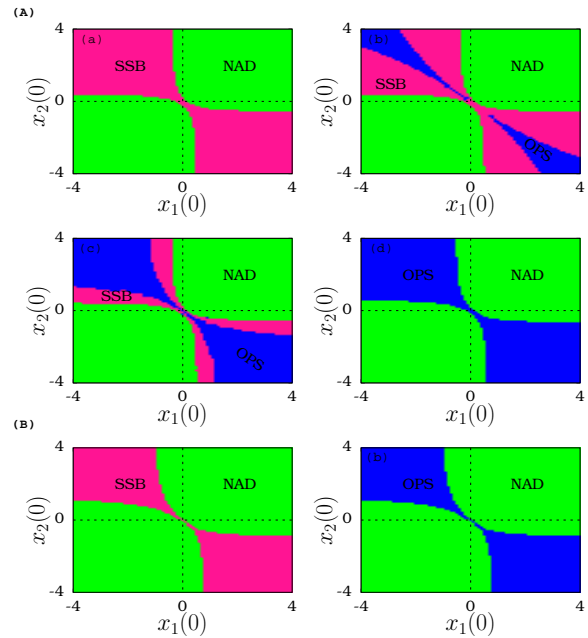}
			\caption{ Basin of attraction obtained by  fixing $y_1(0) = -0.45$ and $y_2(0) = 0.5$ while varying $x_1(0)$ and $x_2(0)$.  Figs.~(A)  are plotted along the line $L_1$ in Fig.~\ref{glob}(b) at $\omega=0.5$, for  coupling strengths  (a) $\epsilon=0.50$, (b) $\epsilon=0.507$,  (c) $\epsilon=0.509$ and (d) $\epsilon=0.52$.   Figs.~(B)  are pictured along the line $L_2$ at $\omega=0.43$ for   (a) $\epsilon=0.576$ and (b) $\epsilon=0.577$.  Other parameters: $\lambda=1.0$ and $a=1.0$.  }
			\label{basin}
			\vspace{-0.4cm}
		\end{figure} 
	
	  In order to understand the multistabilities among the dynamical states  and transition route to OPS in the SSB region,  we portray the basin of attraction in Fig.~\ref{basin}  by fixing $y_1(0) = -0.45$ and $y_2(0)=0.5$ and by varying $x_1(0)$ and $x_2(0)$.   Fig.~\ref{basin}(A): (a)-(d) are plotted for the dashed line along  $L_1$ in Fig.~\ref{glob}(b) at $\omega=0.5$.  At coupling strength $\epsilon=0.5$, we find  the coexistence of SSB and NAD states in the basin of attraction, that is  SSB state for some initial conditions and  NAD state for other initial conditions (see Fig \ref{basin}A(a)).  Here, we notice that the   symmetric initial conditions  (i.e.  $x_1(0)$ and $x_2(0)$ both are either positive or negative)  lead to NAD whereas  asymmetric initial states give rise to SSB states (i.e. in the first and third quadrants where either  $x_1(0)$ or  $x_2(0)$  is positive while the other one is negative). On  increasing the coupling strength  $\epsilon=0.507$,  the emergence of OPS branch inside the SSB region  is observed in   the basin of attraction. Here, the coexistence of  NAD, SSB and OPS states are noticed  with respect to distinct initial conditions. On further increment of $\epsilon$ to 0.509,  the basin of attraction for OPS state is increased and the corresponding SSB basin is decreased. Increasing of coupling strength further, it is seen that the SSB basin is completely occupied by the OPS basin.  In this way the SSB state transits to OPS state through increasing of OPS basin smoothly as a function of  coupling strength.  The transition is also analyzed along the line $L_2$ in  Fig.~\ref{glob}(b) at $\omega=0.43$ in the SSB region, and the corresponding basin of attraction is depicted in Figs.~\ref{basin} B(a)-B(b).  The SSB state in the first  and third quadrant is  directly replaced by OPS basin  during a  small increment of coupling strength from $\epsilon= 0.576$  to $\epsilon= 0.577$.  From the basin stability analysis, we find that the coexistence of distinct dynamical states in the basin of attraction causes multistability in the parametric space and    we  confirm that two distinct types of transition to OPS from the SSB exist and also coexistence of  SSB-NAD, SSB-OPS-NAD and OPS-NAD states.
	From the basin of attraction,  it is also clear that the symmetric initial states lead only to NAD and IPS states whereas the asymmetric initial states give rise to OPS and SSB states. Further, the effect of feedback factor on stable  SSB, NAD and OD  states are demonstrated in the following section. 
		
	\section{Impact  of feedback factor on spontaneous symmetry breaking (SSB) state and oscillation quenching states}

		\par
\begin{figure}
				\centering
				\hspace{-0.4cm}
				\includegraphics[width=8.00cm]{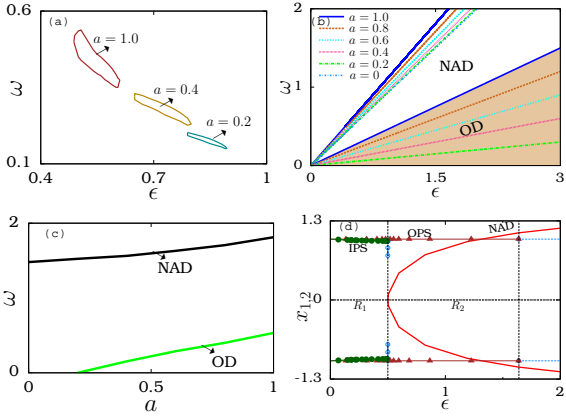}
				\vspace{-0.3cm}
				\caption{(a) Regions of symmetry breaking oscillations in ($\epsilon, \omega$) space  with respect to three distinct values of feedback factor $a$. (b) Stable boundaries  of NAD and OD region in  ($\epsilon, \omega$) space  as  a function of feedback factor $a$. (c) maximum stability region of NAD and OD at $\epsilon=0.5$ with respect to $a$. (d) Bifurcation diagram for $a=0$, $\omega=0.5$.  Here $\lambda$ is fixed as $1$.}
				\label{sbo}
					\vspace{-0.5cm}
			\end{figure}

 To exemplify the effect of feedback factor on SSB we have plotted the regions of   SSB  for three different  values of feedback factors, namely $a=1.0$, $a=0.4$ and $a=0.2$ in the ($\epsilon,\omega$) space in  Fig.  \ref{sbo}(a).  It is evident that there is a decreasing of  SSB regions with decreasing feedback factor  $a$.  From   Fig. \ref{sbo}(a), it is  observed that the  SSB region is relatively  large  for $a=1$ compared to the SSB regions with $a=0.4$ and $a=0.2$.  Decreasing the feedback factor  decreases the SSB region and finally there will be no symmetry breaking oscillations for $a=0$ which completely suppresses the SSB states.  Additionally the   impact of feedback factor is also analyzed in NAD and OD regions.  The shrinking of oscillation death is evident from  Fig. \ref{sbo}(b).  The shaded region in Fig.~\ref{sbo}(b) denotes the OD region for $a=1$   and  distinct lines  represent   different  boundaries of NAD and OD for different values of feedback factor. Decreasing the feedback factor only shrinks the oscillation death state  but not  the nontrivial AD state which sustains even in the absence of feedback factor.   Fig. \ref{sbo}(c) is plotted for $\omega$  as a function of feedback factor $a$.  With respect to  the feedback factor the  boundary of the OD region attains null values of  $\omega$  but  the nontrivial AD states take nonzero values of $\omega$ which confirms the retaining of  AD regions. Furthermore to confirm the observed results, the bifurcation diagram is plotted at $a=0$ in Fig. \ref{sbo}(d) which clearly depicts the emergence of NAD even in the absence of  the feedback factor.     It is observed that the oscillatory region is increased compared to the  presence of feedback factor.    Finally the system shows stable NAD at large coupling strengths and it confirms that the OD is  state entirely suppressed. From the above  results we conclude that the the feedback completely suppresses the SSB and OD regions only but it sustains the NAD region. 
	
	The stability of the observed in-phase and  out-phase oscillations, NAD and OD states are detailed in the next section. 
	
	\section{\label{antica} Stability of distinct dynamical states}
	\par {In order to study the stability  of the observed  dynamical states, we first rewrite  Eq. (\ref{model2})  by separating the symmetric ($z_+$) and anti-symmetric ($z_-$) variables,} where $z_+={(z_1+z_2)}/{2}$ and $z_-={(z_1-z_2)}/{2}$. 
	Eq. (\ref{model2}) in terms of these new variables is given by 
	\begin{subequations}
	\begin{eqnarray}
	\dot{z}_+&=&\frac{1}{2}\left(f(z_++z_-)+f(z_+-z_-)\right)+i \epsilon \, \mathrm{Im}(z_+) \nonumber\\
  && \qquad \qquad \qquad \qquad \qquad \qquad -\epsilon a \,\mathrm{Re}(z_+), \\
	\dot{z}_-&=&\frac{1}{2}\left(f(z_++z_-)-f(z_+-z_-)\right)- \epsilon a\, \mathrm{Re}(z_-).  
	\end{eqnarray} 
	\end{subequations}
\par  In the in-phase subspace, $Z_+=\{(z_+, z_-)|z_- \equiv 0\}$, and in the anti-symmetric subspace, $Z_-=\{(z_+, z_-)|z_+ \equiv 0\}$.  Thus in the symmetric and anti-symmetric subspaces, the dynamical equations can be reduced, respectively, to 
	\begin{eqnarray}
	\dot{z}_+=f(z_+)+i\epsilon\, \mathrm{Im}(z_+)-  \epsilon  a\, \mathrm{Re}(z_+) , \; \; \dot{z}_-=0,
	\label{symsub}
	\end{eqnarray}
	\vspace{-0.1cm}
	and
		\begin{eqnarray}
		\dot{z}_-=f(z_-) - \epsilon{ a} \,\mathrm{Re}(z_-) , \;\; \dot{z}_+=0. 
		\label{ansub}
		\end{eqnarray}
	We find  the explicit expressions for the different oscillatory states, steady states and their stabilities in the following.\vspace{0.5cm}
	{\it (a) Dynamical states in symmetric subspace:} The solution of the corresponding dynamical equation in the symmetric subspace Eq. (\ref{symsub}) is found to be expressed as 
	\begin{eqnarray}
	x(t)&=& \frac{{e^{\frac{\bar{\lambda}}{2}t}} \cos[\delta-\frac{1}{2} \psi t]}{(C-\frac{1}{2\bar{\omega}^2}e^{\bar{\lambda} t }[U_0+U_1 \cos \psi t +U_2 \sin \psi t])^{\frac{1}{2}}}, \nonumber \\
	y(t)&=&\frac{1}{2\bar{\omega}}\left(-{\epsilon   a}+\psi \, \mathrm{\tan} (\frac{1}{2}\psi\,t -\delta) \right) x(t),
	\label{eq6}
	\end{eqnarray}
	where $\psi=\sqrt{4\omega\bar{\omega}-{  \epsilon}^2 a^2 }$, $\bar{\lambda} = 2\lambda -\epsilon a$, $\beta_1=-\lambda((2+  a^2)\epsilon-2\omega)+  a((1+  a^2)\epsilon^2+\epsilon\omega-2\omega^2)$, $\beta_2=( -a \lambda+\epsilon(1+a^2)-\omega )\psi $ and $\bar{\omega}= \omega-\epsilon$. $C$ and $\delta$ are integration constants.  The other constants $U_0$, $U_1$ and $U_2$ are
	\begin{align}
	&U_0= \frac{2(\epsilon-2\omega)\bar{\omega}}{\bar{\lambda}}, \quad U_1= \frac{\epsilon(\beta_1 \cos 2\delta + \beta_2 sin 2 \delta)}{2(\lambda^2-  \lambda\epsilon a +\omega\bar{\omega})}, \nonumber \\ \quad & U_2= \frac{\epsilon(\beta_1 \sin 2\delta - \beta_2 cos 2 \delta)}{2(\lambda^2-  \lambda\epsilon a+\omega\bar{\omega})}.
	\label{Q_0Q_1Q_2}
	\end{align}
	Further, we  can find that  the solution  of  Eq. (\ref{eq6}) will be periodic when $4\omega\bar{\omega} > {\epsilon}^2  a^2$.  In this case, we can write the state variables of  $x_k(t)$ and $y_k(t)$, $k=1,2$, in the asymptotic limit ($t \rightarrow \infty$) as
	\begin{small}
		\begin{align}
		&x_1(t)=\frac{\cos (\delta-\frac{1}{2}\psi \,t)}{\left(U_0+U_1 \cos \psi t +U_2 \sin \psi t\right)^{\frac{1}{2}}}, \nonumber \\
		&y_1(t)=\frac{-1}{2\bar{\omega}}\frac{({\epsilon  a} \cos (\delta-\frac{1}{2}\psi \,t) \, +\psi \,sin (\delta-\frac{1}{2}\psi \,t))}{\left(U_0+U_1 \cos \psi t +U_2 \sin \psi t\right)^{\frac{1}{2}}},
		\label{anosc}
		\end{align}
	\end{small}
	with $x_2=x_1$ and $y_2=y_1$.  	Using the Floquet theory and by integrating the above equations from $t=0$  to $t=T=\frac{2\pi}{\sqrt{4\omega\bar{\omega} - {\epsilon}^2  a^2}}$, we determine the Floquet multipliers $\rho_i\, (i=1,...,4)$ and we conclude that when all the four eigen values lie within the unit circle then the corresponding periodic orbit is stable.   Using  the Floquet multipliers we have obtained the boundary for in-phase synchronized oscillations. 
	\par Whenever $4\omega\bar{\omega}<{\epsilon}^2  a^2$, the solution in Eq. (\ref{eq6}) tends to a homogeneous steady state which is given by 
	\begin{align}
	x_1^* &=  \mp \frac{1}{\sqrt{2}} \bigg(\frac{1}{\hat{a}\epsilon}(-\epsilon a^2 \beta_3+ \bar{\omega}(2\lambda+\psi_1)-a( \beta_4+\lambda \psi_1))\bigg)^\frac{1}{2}, \nonumber \\
	y_1^* &= \mp \frac{1}{2 \bar{\omega}}(\epsilon a + \psi_1)x_1^*,
	\label{anod}
	\end{align}
	with $x_2^*=x_1^*$ and $y_2^*=y_1^*$. Other constants are  $ \psi_1= \sqrt{\epsilon^2a^2-4\bar{\omega}\omega}$, $\hat{a}=(1+a^2)$,  $\beta_3=(\epsilon a +\lambda +\psi)$,  $\beta_4=(\epsilon^2+\epsilon\omega-2\omega^2)$.  The eigen values corresponding to Eq.~(\ref{anod}) can be expressed as
	\begin{eqnarray}
	\mu_{1,2}=\frac{1}{2}\, (-\bar{\lambda}-2\psi_1\mp \sqrt{\bar{\lambda}^2}), \,\, \mu_{3,4}= \frac{1}{2} \bigg(- \bar{\lambda}-2\psi_1 \nonumber \\
	  \nonumber \\ 
	\mp  \quad  \sqrt{\frac{1}{\hat{a}}(\beta_5-4a\omega(2\lambda+\psi_1)+4(\beta_6+\lambda\psi_1)) }\bigg),
	\label{odeig}
	\end{eqnarray}
	where $\beta_5=a^3\epsilon(\epsilon a - 4\lambda)+a^2(\epsilon^2+4\lambda^2)$ and $\beta_6= \lambda^2+(\epsilon-2\omega)\omega$. The stable region of NAD emerges when $\omega< \frac{1}{2}(\epsilon+\sqrt{\epsilon^2+a^2\epsilon^2})$. It is also clear that when varying the coupling strength $\epsilon$, there exists a direct transition from   stable IPS to NAD state.  The analytical boundaries for  IPS and NAD states get well fitted  with the numerically obtained  boundaries (see  Fig.~\ref{glob}). \\

	{\it  (b) Dynamical states in the anti-symmetric subspace:} In this section, we have also  found the solution for the  dynamical equation in the anti-symmetric subspace, Eq.~(\ref{ansub}), which can be written as
	\begin{align} 
	\vspace{-0.3cm}
	&x(t)=\frac{e^{\frac{\bar{\lambda}}{2}t } \cos(\delta-\frac{\theta}{2}t)}{\left(C+e^{\bar{\lambda}t}(V_0+V_1 \cos (2\delta- \theta t)+V_2 \sin(2\delta-\theta t))\right)^{\frac{1}{2}}}, \nonumber \\
	&y(t)=\frac{1}{2\omega}\left(-{\epsilon   a}+ \theta \, \mathrm{tan}(\frac{1}{2}\theta t-\delta)\right) x(t),
	\label{exact}
	\end{align}
	where $\theta=\sqrt{4\omega^2-{\epsilon}^2  a^2}$. $C$ and $\delta$ are integration constants.  The other constants $V_0$, $V_1$ and $V_2$ are
	\begin{align}
	&V_0=\frac{2}{2\lambda-  a\epsilon}, \quad V_1=\frac{  \epsilon a(  \lambda\epsilon a-  \epsilon^2 a^2 +2\omega^2)}{4\omega^2(\lambda^2-  \lambda \epsilon a +\omega^2)}, \nonumber \\ \quad &V_2=\frac{  \epsilon a (\lambda-  \epsilon a)\theta}{4\omega^2(\lambda^2-  \lambda \epsilon a+\omega^2)}.
	\label{Q_0Q_1Q_2}
	\end{align}
\par The solution given by Eq.~(\ref{exact}) is periodic when $4\omega^2 > {\epsilon}^2  a^2$.  In the asymptotic limit ($t \rightarrow \infty$),  the state variables $x_k$ and $y_k$, $k=1,2,$  can be  written as
	\begin{small}
		\begin{align}
		&x_1(t)=\frac{\cos (\delta-\frac{1}{2})\theta t}{\left(V_0+V_1 \cos (2\delta-\theta t)-V_2 \sin(2\delta-\theta t)\right)^{\frac{1}{2}}}, \nonumber \\
		&y_1(t)=\frac{-1}{\omega}\frac{({\epsilon  a} \cos (\delta-\frac{1}{2}\theta t)+\theta \sin (\delta-\frac{1}{2}\theta t))}{\left(V_0+V_1 \cos (2\delta -\theta\,t)+V_2 \sin(2\delta-\theta\,t)\right)^\frac{1}{2}},
		\label{anosc}
		\end{align}
	\end{small}
	with $x_2=-x_1$ and $y_2=-y_1$.  {Note that the  solution given in Eq.~(\ref{anosc}) is periodic with respect to the period t=T= $\frac{2 \pi}{\sqrt{4\omega^2-{\epsilon}^2  a^2}}$.  Then using the Floquet multipliers, we have figured out the  boundary of stable OPS state.  }
	\par Further, the solution in Eq.~(\ref{exact})  tends toward a steady state when $4\omega^2<{\epsilon}^2  a^2$, which can be expressed as 
	\begin{align}
	x_1^*&= \mp \frac{1}{\sqrt{2}} (\lambda-\epsilon a +\frac{1}{\epsilon a}(2 \omega^2-\lambda \theta_1)+\theta_1)^\frac{1}{2}, \nonumber \\
	y_1^* &=  \pm \frac{1}{2 \omega} (\epsilon a+\theta_1)x_1^*
	\label{eq16}
	\end{align}
	with $x_2^*=-x_1^*$ and $y_2^*=-y_1^*$.   $\theta_1= \sqrt{\epsilon^2 a^2-4\omega^2}$, 	we have studied the stability of such anti-symmetric steady states Eq. (\ref{eq16})  too and found that the eigenvalues corresponding to the states are given by
	\begin{eqnarray}
	\mu_{1,2}= \frac{1}{2} \,(-\bar{\lambda}-2 \theta_1 \mp \sqrt{\bar{\lambda}^2}), \,\,\, \mu_{3,4}=\frac{1}{2}\, (-\bar{\lambda}-2 \theta_1 \nonumber \\
	 \mp  \sqrt{\frac{1}{a}(\epsilon a^2(\epsilon a - 4 \lambda)+4(\lambda^2-2\epsilon\omega)+4\omega(2\lambda+\theta_1)}), 
	\label{odeig}
	\end{eqnarray}
	where $\bar{\lambda}=(\lambda-\epsilon a)$. From the above eigen values we have found the stability condition for oscillation death state. When $\omega< \frac{a\epsilon}{2}$, the  stable OD region emerges. The obtained analytical stability curves of OPS and OD regions exactly matches with the numerically obtained  boundaries (see  Fig.~\ref{glob}).  From this,  we have verified the stability curves of observed dynamical states of IPS, OPS, NAD and OD regions analytically too.

	\section{Conclusion} 
	We have investigated the dynamical behavior in a pair of conjugately coupled Stuart-Landau oscillators. Interestingly,  we found  the emergence of spontaneous symmetry breaking  due to breaking of permutational symmetries.  We also  noticed the coexistence of distinct dynamical states such as SSB-NAD, NAD-OD, and IPS-OPS states in the parametric space. Through a basin of attraction study,  the dynamical transition to OPS state was illustrated in the symmetry breaking region.     In particular, we report that  the transition to OPS was found as smooth  and sudden transition  as a function of coupling strength.  Further we have analyzed the effect of feedback factor  on the SSB, NAD and OD states.   We found the decreasing feedback factor completely shrinks the symmetry breaking oscillation state and oscillation death state but it does not completely suppress the nontrivial AD state.  Finally, we have also deduced the  analytical expressions for  the observed in-phase and  out-of-phase oscillations as well as nontrivial amplitude death and oscillation death states.  We also find that the feedback factor entirely shrinks oscillation death states  and nontrivial AD states sustained even in the absence of feedback factor. The obtained results are having good agreement with laser applications where we need to suppress unwanted fluctuations and attain constant output. We believe that the present study will shed	light on the dynamics of symmetry breaking states and control of such dynamical behavior, which will be helpful in  many physical and biological systems.
	\vspace{-0.5cm}
	\acknowledgments
    KP and VKC are supported by SERB-DST Fast Track scheme for young scientists under Grant No. YSS/2014/000175.   KS sincerely thanks the CSIR for fellowship under SRF Scheme (09/1095(0037)/18-EMR-I). The work of M.L.
    is supported by a DST- SERB Distinguished Fellowship
    program. 


\vspace{-0.5cm}
	\begin{thebibliography}{0}
		\bibitem{book1}
		\Name { Pikovsky A. S.,  Rosenblum M. G., and  Kurths J.}
		\Book{Synchronization. A Universal Concept in Nonlinear Sciences}
		\Publ{Cambridge University Press, Cambridge, UK}
		\Year{2001}
		
		\bibitem{syn1}
		\Name { Tinsley M. R., Nkomo  S., and  Showalter K.}
		\REVIEW{Nat. Phys.}{8} {2012}{662}.
		
		\bibitem{syn2}
		\Name 	{ Williams C. R. S.,  Murphy T. E.,  Roy R.,  Sorrentino F., 
			Dahms T., and  Sch\"oll E.}
		\REVIEW{Phys. Rev. Lett.} {110} {2013}{064104}.
		
		\bibitem{syn3}
		\Name{ Pecora L. M.,  Sorrentino F.,  Hagerstrom A. M.,  Murphy T. E.,
			and  Roy R.}
		\REVIEW{Nat. Commun.} {5}{2014} {4079}.
		
		\bibitem{os1}
		\Name{ Aronson D. G.,  Ermentrout G. B., and  Kopell N.}
		\REVIEW{Physica D} {41}{1990}{403}.
		
		\bibitem{os2}
		\Name{Mirollo R. E.  and  Strogatz S. H.}
		\REVIEW{J. Stat. Phys.}{60} {1990}{245}.
		
		\bibitem{ADOD}
		\Name{  Banerjee T. and Ghosh  D.}
		\REVIEW{Phys. Rev. E}{89}{2014} {062902}; {\textit Phys. Rev. E} {\bf 89} {(2014)} {052912}.
		
		\bibitem{AD-OD}
		\Name{ Zou W., Senthilkumar	D. V.,  Duan J., and  Kurths J.}
		\REVIEW{Phys. Rev. E}{90}{2014}{032906}. 
		
		\bibitem{ap1}
		\Name{ Matthews P. C. and  Strogatz S. H.}
		\REVIEW{Phys. Rev. Lett.}{65}{1990} {1701}.
		
		\bibitem{ap2}
		\Name{ Dolnik M. and Epstein  I. R.}
		\REVIEW{Phys. Rev. E} {54}{1996}{3361}. 
		
		\bibitem{ap3}
		\Name { Hohl A., Gavrielides A.,  Erneux T., and  Kovanis V.}
		\REVIEW{Phys. Rev. Lett.} {78}{1997}{4745}.   
		
		\bibitem{ssb_turing}
		\Name{ Sawai S., Maeda Y., and  Sawada Y.}
		\REVIEW{Phys. Rev. Lett.}{85} {2000}2212.	
			
		\bibitem{turing}
		\Name{ Pismen L. M.}
		\REVIEW {J. Chem. Phys.} {101}{1994}{3135}.
		
		\bibitem{comp}
		\Name { Singh R.,  Menon S. N., and  Sinha S.} 
		\REVIEW{Sci. Rep.}{ 6}{2016} {22074}.
	
		\bibitem{sbo}
		\Name{ Sathiyadevi K.,  Karthiga S.,  Chandrasekar V. K,  Senthilkumar D. V,  Lakshmanan M.}
		\REVIEW{Phys. Rev. E}{95}{2017}	{042301}.  
		
		\bibitem{od_chem_osc}
		\Name{ Hynne F., and  Graae Soerensen P.}
		\REVIEW{J. Phys. Chem.}{91} {1987}{6573}.
		
		
		\bibitem{revi2}
		\Name{ Koseska A.,  Volkov E., Kurths  J.}
		\REVIEW{Phys. Rep.}{ 531}{2013}{173}.
		
		\bibitem{dvs_nature}
		\Name { Zou W.,  Senthilkumar D. V.,Nagao  R.,  Kiss I. Z.,  Tang Y.,  Koseska A.,
			Duan  J., and Kurths J. }
		\REVIEW {Nat. Commun.}{6}{2015}{ 7709}.
		
		\bibitem{liu_od} 
		\Name{Chen J., ,  Liu W.,  Zhu Y.,  and  Xiao J., } 
		\REVIEW{Euro. Phys. Lett.}{115}{2016}{20011}
		
		\bibitem{saxena}
		\Name{ Saxena G.,  Prasad A. and  Ramaswamy R.}
		\REVIEW{Phys. Rep.}{521}{2012} {205}.
		
		\bibitem{conj_td}
		\Name{ Karnatak R., Ramaswamy  R. ,  and Prasad  A.  }
		\REVIEW{Phys. Rev. E}{76}{ 2007 } {035201}.
		
		\bibitem{td0}
		\Name{Ramana Reddy D. V., Sen A. and Johnston G. L.}
		\REVIEW{Phys. Rev. Lett.}{80}{1998}{5109};
		\REVIEW{Phys. Rev. Lett.}{85} {2000} {3381};
		\bibitem{td1}
		\Name{ Strogatz S. H.} 
		\REVIEW{Nature, London} {394}{1998}{316}. 
		
		\bibitem{td2}
		\Name{Prasad A.}
		\REVIEW{Phys. Rev. E}{72}{2005} {056204}; 
		\Name{Vicente R.,  Tang S.,  Mulet J., Mirasso  C. R., and Liu J. M.}
		\REVIEW{Phys. Rev. E}{73}{2006}{047201}.
		
		\bibitem{td3}
		
		\Name{  Bera B.K.,  Hens C. R.,  Ghosh D.}
		\REVIEW{ Phys. Lett. A}{380}{2016}{2366-2373}. 
		\bibitem{fm}
		\Name{Mirollo R. E. and Strogatz S. H.}
		\REVIEW{ J. Stat. Phys.} {60} {1990} {245}.
		
		\bibitem{dynamic} 
		\Name{ Konishi K. }
		\REVIEW {Phys. Rev. E}{68}{2003}{067202}. 
		
		\bibitem{dir_indi}	
		\Name{Majhia S., Bera B. K., Bhowmick S. K. Ghosh. D.}	
		\REVIEW	{Phys. Lett. A} {\bf 380}{2016} {3617-3624}.
		
		
		\bibitem{laser_ad}
		\Name{  Kim M.Y.,  Roy R.,  Aron J.L.,  Carr T.W.,  Schwartz I.B.}
		\REVIEW{Phys. Rev. Lett.}{94}{2005}{088101}.
		
		\bibitem{laser_ad1}
		 \Name{ Prasad A.,  Lai Y. C.,  Gavrielides A., and Kovanis V. }
		 \REVIEW{Phys. Lett. A} {318}{2003}{71}.
		 
		
		\bibitem{neu_ad}
		\Name{ Zhai Y., Kiss  I.Z., Hudson  J.L.}
		\REVIEW{Phys. Rev. E}{69}{2004}{026208}.
		
		
		\bibitem{od_prl}
		\Name{ Koseska A.,  Volkov E., Kurths  J.}
		\REVIEW { Phys. Rev. Lett.} { 111}{2013} {024103}.
		
		\bibitem{delay_od1}
		\Name{Zou W.,  Senthilkumar D. V.,  Tang Y., and  Kurths J.}
		\REVIEW{Phys. Rev. E} {86}{2012} {036210}. 
		
		\bibitem{delay_od2}
		\Name{ Zou W.,  Senthilkumar D. V.,  Duan J., and  Kurths J.}
		\REVIEW{Phys. Rev. E}{ 90}{2014} {032906}. 
		
		\bibitem{conj1}
		\Name{ Han W.,  Cheng H., Dai  Q., Li H.,  Ju P.,
			 Yang J.}
		\REVIEW{Commun Nonlinear Sci. Numer. Simulat.}{39}{ 2016 } {73}.
			
			
		\bibitem{rep_od}
		\Name{ Hens C. R.,  Olusola O. I.,  Pal P., and  Dana S. K.} 
		\REVIEW{Phys. Rev. E}{88}{2013}{034902}.
		
		\bibitem{sgo1}
		\Name{ Koseska A., Volkov E.,  Zaikin A., and  Kurths J.}
		\REVIEW{Phys. Rev. E}{75}{2007} {031916}. 
		
		\bibitem{sgo2}
		\Name{ Ullner E., Zaikin A.,  Volkov E. I., and  García-Ojalvo J.}
		\REVIEW{Phys. Rev. Lett.}{99}{2007}{ 148103}.;  
		\Name{ Ullner E., Koseska  A.,  Kurths J.,  Volkov E.,  Kantz H., and  García-Ojalvo J.,}
		\REVIEW{Phys. Rev. E}{78} {2008}{ 031904}. 
		
		\bibitem{stem}
		\Name{ Goto Y. and  Kaneko K.}
		\REVIEW{Phys. Rev. E}{88}{2013} {032718}.
			
		\bibitem{cardio}
		\Name{ Suárez-Vargas J. J.,  González J. A.,  Stefanovska A., and
		McClintock P. V. E. }
	   \REVIEW{ Euro. Phys. Lett.}{85}{2009}{38008}.
		
			\bibitem{liu_revive}
			\Name{Deng T., Liu W.,  Xiao J., and   Kurth J.}
			\REVIEW{Choas}{26}{2016}{094813}
			
			
		
		\bibitem{vkc}
		\Name{ Chandrasekar V. K.,  Karthiga S., and  Lakshmanan M.}
		\REVIEW {Phys. Rev. E} {92}{2015} {012903}.
		\bibitem{AD_revok}
		
			\Name{ Majhi S., and  Ghosh D.}
			\REVIEW{Euro. Phys. Lett.}{118}{2017} {40002}.
		
		\bibitem{conj_scl}
		\Name{	 Kim M.Y.,  Roy R.,  Aron J.L.,  Carr T.W.,  Schwartz I.B.}
		\REVIEW {Phys. Rev. Lett.}{94}{2005}{088101}.
		
		\bibitem{prey}
		\Name{ Karnatak R.,  Ramaswamy R., Feudel  U.}
		\REVIEW {Chaos Solitons Fract.}{68}{2014}{48}.
	
		\bibitem{conj(2017)}
		\Name 	{ Zhao N.,  Sun Z.,  Yang X. and  Xu. W.}
		\REVIEW{Euro. Phys. Lett.}{118}{2017} {30005}.
		
		\bibitem{conj_vdp(2018)}	
		\Name{Zhao N., Sun Z., Yang X., and Xu. W.}
		\REVIEW {Phys. Rev. E} {97}{2018} {062203}.
		 \bibitem{basin}
		
		 	\Name{Rakshit S.,  Bera B. K., Majhi S., Hens C.R.,   and  Ghosh D.}
		 	\REVIEW{Scientific Reports}{7}{2017}{45909}.
		 
		
		\bibitem{model_stu}
		\Name  {Schmidt L.,   Sch\"onleber K.,  Krischer K., and  Garc\'ia-Morales V.}
		\REVIEW {Chaos} {24}{2014} {013102}.
		
		\bibitem{lambda}
		
    	\Name{Kundu S.,   Majhi S.,  Karmakar P.,   Ghosh D., 	and  Rakshit B.}
    	\REVIEW{Euro. Phys. Lett.}{123}{2018} {30001}.
    
    
     
	\end{thebibliography}
\end{document}